# Enhanced Infrared Emission by Thermally Switching the Excitation of Magnetic Polariton with Scalable Microstructured VO$_2$ Metasurfaces

*Linshuang Long, Sydney Taylor, and Liping Wang*[*]

School for Engineering of Matter, Transport & Energy, Arizona State University, Tempe, Arizona, USA 85287

**ABSTRACT** Dynamic radiative cooling attracts fast-increasing interest due to its adaptability to changing environment and promises for more energy-savings than the static counterpart. Here we demonstrate enhanced infrared emission by thermally switching the excitation of magnetic polariton with microstructured vanadium dioxide (VO$_2$) metasurfaces fabricated via scalable and etch-free processes. Temperature-dependent infrared spectroscopy clearly shows that the spectral emittance of fabricated tunable metasurfaces at wavelengths from 2 to 6 μm is significantly enhanced when heated beyond its phase transition temperature, where the magnetic polariton is excited with metallic VO$_2$. The tunable emittance spectra are also demonstrated to be insensitive to incidence and polarization angles such that the VO$_2$ metasurface can be treated as a diffuse infrared emitter. Numerical optical simulation and analytical inductance-capacitance model elucidate the suppression or excitation of magnetic polariton with insulating or metallic VO$_2$ upon phase transition. The effect of enhanced thermal emission with the tunable VO$_2$ metasurface is experimentally demonstrated with a thermal vacuum test. For the same heating power of 0.2 W, the steady-state temperature of the tunable VO$_2$ metasurface emitter after phase transition is found to be 20°C lower than that of a reference V$_2$O$_5$ emitter whose static spectral emittance is almost the same as that of the VO$_2$ metasurface before phase transition. The radiative thermal conductance for the tunable metasurface emitter is found to be 3.96 W/m$^2$K with metallic VO$_2$ at higher temperatures and 0.68 W/m$^2$K with insulating VO$_2$ at lower temperatures, clearly demonstrating almost six-fold enhancement in radiative heat dissipation.

## Introduction

Thermal management is crucial for many applications, such as buildings,[1] solar cells,[2] and electronics.[3] Radiation-based thermal management is a promising addition or alternative to conventional methods.[4-5] In particular, radiation heat transfer is effective in spacecraft thermal control[6] and terrestrial radiative cooling by dumping heat from earth to the cold space through the atmospheric window.[7] Performance of radiative heat dissipation heavily depends on the surface radiative properties, which are usually static or a weak function of temperature. Recently, dynamic control of radiative properties has attracted increasing interest because of the adaptability to the changing environment and the promise for more energy savings. For example, Ye et al. suggested that the emissivity of a window should be low in winter and high in summer for energy efficiency in both seasons.[8] The surface of a spacecraft is expected to be of high emittance to enhance cooling when the spacecraft temperature is high, while the emittance should be low when the temperature is low to reduce heat loss.

One key to dynamic radiative thermal control lies in materials whose radiative properties are tunable in response to external stimuli. Among those chromogenic materials, vanadium dioxide ($VO_2$) attracts lots of research attentions as it could undergo a reversible phase transition from an insulating state to a metallic state when its temperature exceeds 68°C.[9] This structural change causes a significant change in optical properties, which makes $VO_2$ attractive for dynamic radiation control. By employing nanostructured $VO_2$, reconfigurable metadevices with tunable absorption were experimentally realized.[10] Radiative thermal runaway was experimentally investigated with a $VO_2$-based thermal emitter.[11] A tunable hyperbolic metasurface comprised of a hexagonal boron nitride (hBN) and $VO_2$ was studied.[12] Kocer et al. demonstrated a tunable absorber based on a hybrid gold-$VO_2$ nanostructure.[13] A multilayer structure including a $VO_2$ film was fabricated to achieve a thin-film optical diode.[14]

One major challenge in utilizing $VO_2$ for dynamic thermal control is that, the emittance of pristine $VO_2$ film decreases as it becomes metallic at high temperatures, which actually desires larger emittance to promote the infrared emission for cooling. To address this property mismatch, nanophotonic and metamaterial structures have been proposed to achieve the desired switch of thermal emittance. $VO_2$-based multilayers have been theoretically designed for active radiative cooling[15-16] and radiative thermostats.[17] A metasurface absorber was theoretically proposed through a combination of $VO_2$ and hBN.[18] Very few studies have experimentally achieved significantly increased emissivity at higher temperatures with $VO_2$-based metamaterials. While Ito et al.[19] showed the tunable infrared emittance of $VO_2$ patches fabricated from etching process, direct experimental demonstration of enhanced emissive power is not reported yet.

In this work, we experimentally demonstrate the augmented infrared emission by thermally switching the excitation of magnetic polariton (MP)[20-21] with scalable microstructured $VO_2$ metasurfaces, whose infrared radiative properties are comprehensively characterized at different wavelengths, temperatures, incident angles, and polarization states. Numerical optical simulation is carried out to validate the unique spectroscopic properties, and the underlying physics of thermally switchable MP with different phases of $VO_2$ is elucidated along with an analytical capacitor-inductor circuitry model. Finally, a thermal vacuum test is conducted to directly demonstrate the quantified performance of enhanced radiative heat dissipation with the tunable $VO_2$ metasurfaces at elevated temperatures.

## Results and discussion

The $VO_2$ metasurface was fabricated by etch-free and scalable patterning processes as respectively illustrated in **Figure 1**(a) and (b) (See Methods for details about sample fabrication). In particular, $VO_2$ metasurfaces were directly oxidized from pre-patterned vanadium (V) microdisks without etching. In conventional methods, $VO_2$ micro- or nano-patterns were formed by selectively etching a pre-deposited $VO_2$ film, which would decrease the conformality[13, 19]. The transformation from V to $VO_2$ was characterized through Raman spectroscopy in Figure 1(c). Pure vanadium has no obvious peak between the Raman shifts of 100 and 1100 cm$^{-1}$, while after the oxidization several characteristic Raman peaks of insulating $VO_2$ appear at 20°C.[22-23] The phase change from insulating to metallic $VO_2$ was further confirmed by a featureless Raman spectrum when the sample was heated at 100°C above the transition temperature of $VO_2$ (~68°C). The height of the V microdisks is about 50 nm before oxidation but doubled to 100 nm after oxidation after $VO_2$ microdisks are formed from profilometry measurements in Figure 1(d). The morphology of the microdisks was examined with scanning electron microscopy (SEM) before and after the oxidation with images respectively shown in Figure 1(e) and 1(f). Clearly, the period ($\Lambda = 1.6$ μm) and diameter ($D = 1.1$ μm) of microdisks do not change and the circular profile was well preserved during the oxidation from V to $VO_2$. In order to achieve large sample areas in cm-scale for radiative thermal tests, translational patterning was implemented during the photolithography process with the stepper as illustrated in Figure 1(b). Multiple $VO_2$ metasurface samples in different pattern sizes from 4 mm to 4 cm squares were successfully fabricated as pictured in Figure 1(g).

The spectral emittance of the $VO_2$ metasurface were measured by a Fourier-transform infrared spectrometer (FTIR) from room temperature up to 120°C at different incident angle $\theta$ and polarization angle $\varphi$ as illustrated in **Figure 2**(a) (See Methods for details about optical characterization). As shown in Figure 2(b), when the temperature increases, the spectral emittance $\varepsilon_\lambda$ measured at near-normal ($\theta = 8°$) gradually increases. Within the wavelength range from 2 to 10 μm, the metasurface shows relatively low emittance and one emission peak around the wavelength of 3 μm when the $VO_2$ is insulating at 25°C. Once heated above the $VO_2$ transition temperature where the insulating $VO_2$ becomes metallic, the tunable metasurface exhibits dramatic changes in spectral emittance mainly due to another major emission peak at a wavelength of $\lambda \approx 5$ μm appears. At temperature of 100°C, the emittance at both peak wavelengths of 3 μm and 5 μm almost reaches unity as a black surface. Clearly, the spectral emittance of the tunable metasurface exhibits a significant variation with temperature, in particular, the emittance is much enhanced at higher temperatures after $VO_2$ transits from insulator to metal. At $\lambda = 5$ μm of the thermally switchable peak, the temperature-dependent emittance changes from 0.2 to 1 from 25°C to 100°C as seen in Figure 2(c), where a typical hysteresis of ~20°C upon $VO_2$ phase transition is observed between the heating and cooling curves.[24-25]

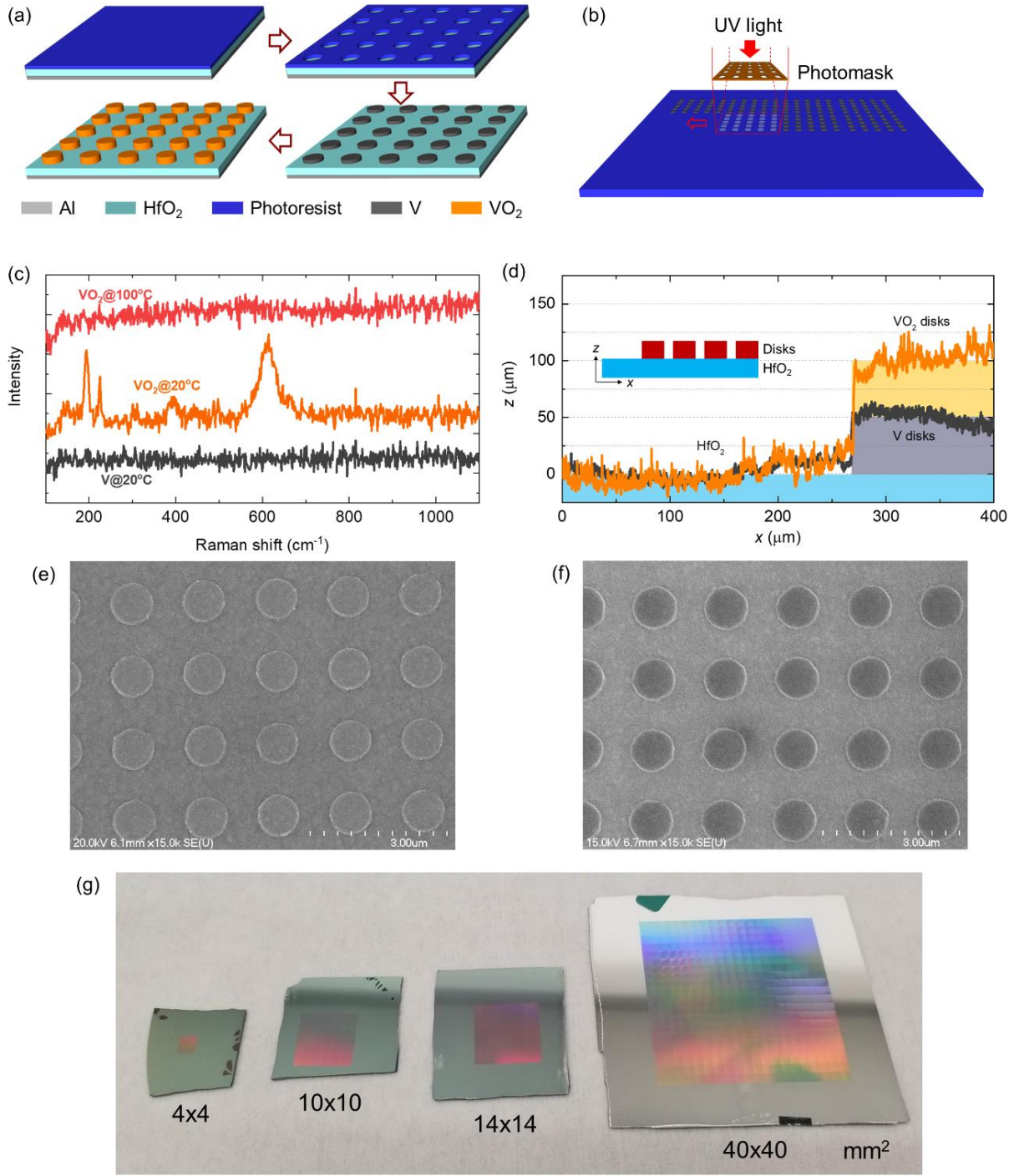

**Figure 1**. Fabrication and characterization of $VO_2$ metasurfaces with etch-free and large- scalable process. (a) Flow chart of the fabrication process. (b) Schematic of producing large-area patterns through stepper. (c) Raman spectrum of V and $VO_2$ at different temperatures. (d) Disk heights from profilometer measurement. SEM images of (e) V and (f) $VO_2$ disks. (g) Photos of samples with different pattern sizes.

For radiative heat dissipation, a diffuse surface is usually preferred to maximize the thermal emission radiated hemispherically to the environment. FTIR measurements at different incident angles $\theta$ and polarization angles $\varphi$ were conducted to investigate the angular and polarization dependency of the tunable metasurface emitter. Figure 2(d) and (e) provide the measured spectral emittance with metallic $VO_2$ (100°C) at different $\theta$ under transverse magnetic (TM, i.e., $\varphi = 90°$) and transverse electric (TE, i.e., $\varphi = 0°$) waves, respectively. Under TM waves, the emittance peak near $\lambda = 3$ μm is sensitive to incident angles, and it splits into several peaks as the incident angle increases. On the other hand, the emittance peak at $\lambda = 5$ μm with metallic $VO_2$ is nearly angle-independent in both magnitude and location for $\theta < 70°$. The diffuse behavior of emission or absorption is one of the typical features of the magnetic polariton (MP) excitation, whose mechanism and characteristics will be elucidated later with numerical simulations. Generally MP can only be excited under TM waves but not TE waves for a 1D structure[26]. Here we overcame this limitation by employing symmetric structures of 2D microdisk arrays in order to maximize the thermal emission in both polarizations. As shown in Figure 2(e) for TE waves, the measured spectral emittance changes little from normal direction to $\theta = 60°$. To better illustrate the polarization independency of the $VO_2$ metasurface, the peak emittance at $\lambda = 5$ μm with metallic $VO_2$ was measured at a fixed incidence angle $\theta = 45°$ for multiple polarization angles $\varphi$ from $0°$ to $180°$. As clearly shown in Figure 2(f), the high emittance is almost constant around 0.95 independent of polarization angles.

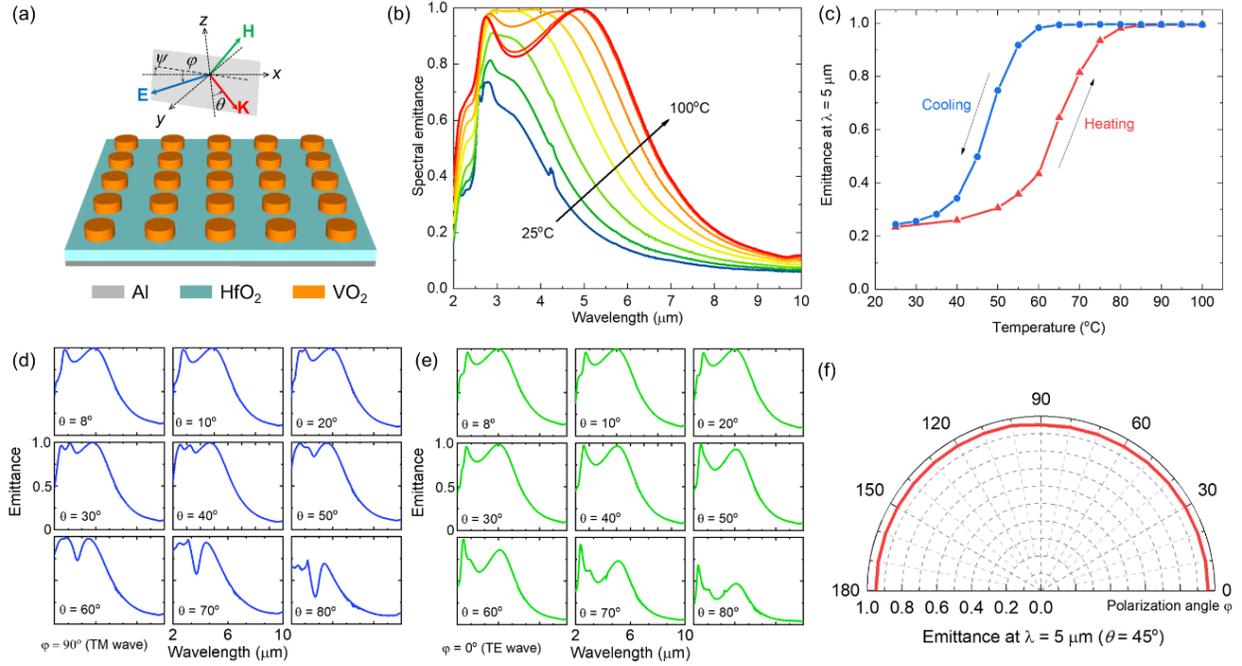

**Figure 2.** FTIR measurement of the as-fabricated metamaterial. (a) Schematic of the directional measurement. (b) FTIR measurement at different temperatures (25, 50, 60, 65, 70, 75, 85, and 100°C) with unpolarized light and near-normal incidence. (c) Emittance at $\lambda = 5$ μm as a function of temperature during the heating and cooling cycles. Incident angle $\theta$ dependent measurement for the metamaterial with metallic $VO_2$ under (d) TM and (e) TE waves. (f) Emittance at $\lambda = 5$ μm as a function of polarization angle $\varphi$ at $\theta = 45°$.

In addition to the optical measurements, full-wave numerical optical simulation is performed to elucidate the underlying physics of the tunable emission (See Methods for details about optical simulation). The simulated spectral emittance under normal incidence is plotted in **Figure 3**(a) for the tunable metasurfaces with insulating and metallic $VO_2$. A similar switchable emittance peak at $\lambda = 5$ μm is observed, which is in good agreement with the measurement. The minor emittance peaks around $\lambda = 3$ μm are also captured by the simulation but much sharper than the measured ones. These minor peaks are associated with surface plasmon polaritons (SPPs) coupled with the period of microdisk arrays via diffraction, whose strong angular dependence could explain the slight difference in the wavelength and bandwidth from the measurement done in a near-normal (8°) direction rather than $\theta = 0°$ in the simulation.[27]

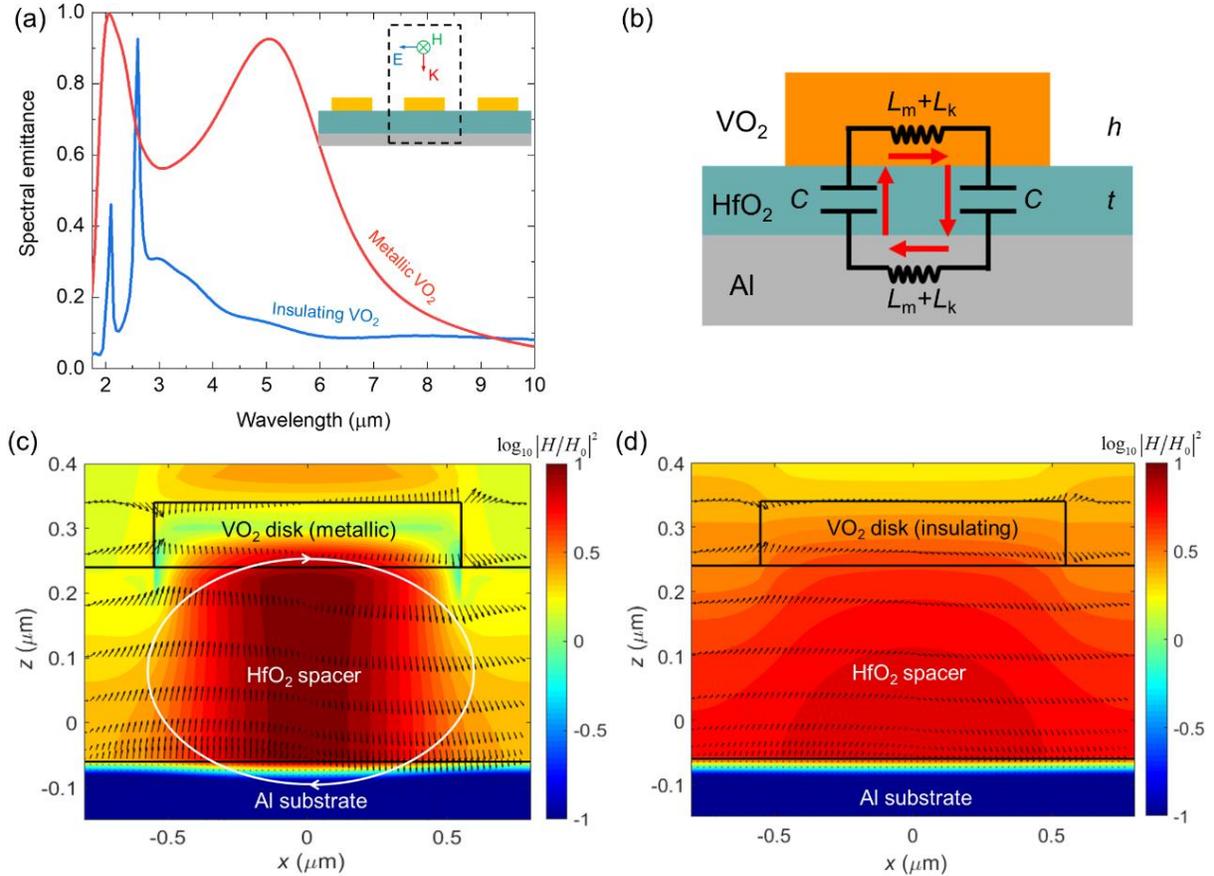

**Figure 3.** Optical simulation of tunable $VO_2$ metasurface emitter. (a) Simulated spectral normal emittance with insulating and metallic $VO_2$. (b) Equivalent LC circuit model for MP excitaton. Electromagnetic field distribution at $\lambda = 5$ μm for (c) metallic and (d) insulating $VO_2$.

The thermally-switchable emittance peak around $\lambda = 5$ μm is associated with magnetic polariton, which can be excited only with metallic $VO_2$. As depicted with an equivalent inductor-capacitor (LC) circuit model in Figure 3(b), the dielectric $HfO_2$ spacer contributes capacitance $C$, and the electromagnetic coupling between the metallic $VO_2$ microdisk and Al substrate introduces mutual inductance $L_m$ and kinetic inductance $L_k$. Magnetic resonance or MP will occur at the resonance wavelength $\lambda_{MP} = 2\pi c_0 \sqrt{(L_m + L_k)C}$ which zeros the total impedance of the LC circuit. Using the previously developed model with expressions for $C$, $L_m$ and $L_k$,[28] the MP resonance wavelength is analytically predicted to be $\lambda_{MP}$ = 4.86 μm, which agrees well with the emittance peak location of $\lambda$ = 5 μm from the optical measurement and simulation. On the other hand, when the VO$_2$ is insulating the mutual and kinetic inductance disappear and the same resonance condition cannot be satisfied. In other words, magnetic resonance is switched "off" such that the thermal emission or absorption from the metasurface is significantly suppressed.

To further confirm that the tunable infrared emittance with different VO$_2$ phases stems from the excitation or suppression of MP, the cross-sectional electromagnetic field distribution across the center of the VO$_2$ microdisk at the peak emittance wavelength of $\lambda$ = 5 μm is plotted in Figure 3(c) for the metallic VO$_2$. The squared magnetic field amplitude normalized to the incidence is represented by contours and the electric field vectors are indicated by arrows. Clearly there is a strong energy confinement within the HfO$_2$ spacer between the metallic VO$_2$ microdisk and the Al substrate along with an electric current loop indicated by the electric field vectors, which signify the excitation of magnetic polariton (MP) inside the structures.[20-21] Undoubtedly, the pronouncedly enhanced infrared emittance spectrum including the unity emittance at $\lambda$ = 5 μm is due to the resonance excitation of MP with metallic VO$_2$. In addition, the spectral absorption or emission peak location associated with MP is known insensitive to incidence angles,[29] which explains the diffuse behavior of the VO$_2$ metasurface emitter observed from the angular optical measurements. When VO$_2$ is at its insulating phase, there is no apparent confinement of the magnetic field in the HfO$_2$ spacer as seen from the electromagnetic field distribution in Figure 3(d), which verifies suppression of MP and the resulting low infrared emittance of the tunable metasurface.

With the demonstrated temperature-dependent tunable infrared emittance, the proposed VO$_2$ metasurface emitter holds great potential for enhancing radiative heat dissipation at high temperatures. A thermal vacuum test was conducted to experimentally assess the emissive power upon the phase change for the tunable $VO_2$ metasurface emitter along with a reference $V_2O_5$ emitter as shown in **Figure 4**(a) (See Methods for details about thermal vacuum test). As shown in Figure 4(b), the total emittance of the tunable $VO_2$ metasurface experiences a steep rise by more than 180% upon phase change with temperature from 40°C ($\varepsilon = 0.08$) to 80°C ($\varepsilon = 0.23$), and finally reaches 0.33 at 160°C. On the other hand, the $V_2O_5$ metasurface emitter, which was fabricated by over-oxidizing the V microdisks and did not undergo the phase transition upon heating, exhibits temperature-independent spectral emittance from the optical measurement shown in the inset. The total emittance of the $V_2O_5$ metasurface is around 0.06 at temperatures below 60°C, which is comparable to the low emittance of tunable metasurface with insulating $VO_2$ before phase transition, and reaches only 0.1 at 160°C. Therefore, the $V_2O_5$ metasurface was used as a static reference emitter here without phase change to manifest the enhanced radiative heat dissipation at higher temperatures from the tunable $VO_2$ metasurface upon phase change.

Steady-state temperatures of both tunable $VO_2$ and static $V_2O_5$ metasurface samples from the thermal vacuum test are presented at varying heating power $q_{in}$ in Figure 4(c), where experimental data is plotted by markers. At the initial heating with power $q_{in}$ up to 0.05 W, both metasurface emitters have almost the same steady-state temperatures from ambient temperature 21°C up to 60°C, at which $VO_2$ phase change started. With the heating power exceeding 0.05 W, temperatures of both samples further increase, while the increment for the tunable $VO_2$ metasurface emitter is appreciably lower than that of the static $V_2O_5$ emitter, indicating better radiative heat dissipation effect, thanks to the escalated infrared emittance by excitation of MP with metallic $VO_2$ at higher temperatures. At the maximal heating power of 0.21 W, the static $V_2O_5$ emitter reaches the highest temperature of 138°C, while the tunable VO$_2$ metasurface emitter is about 20°C cooler, experimentally demonstrating the enhanced radiative cooling effect.

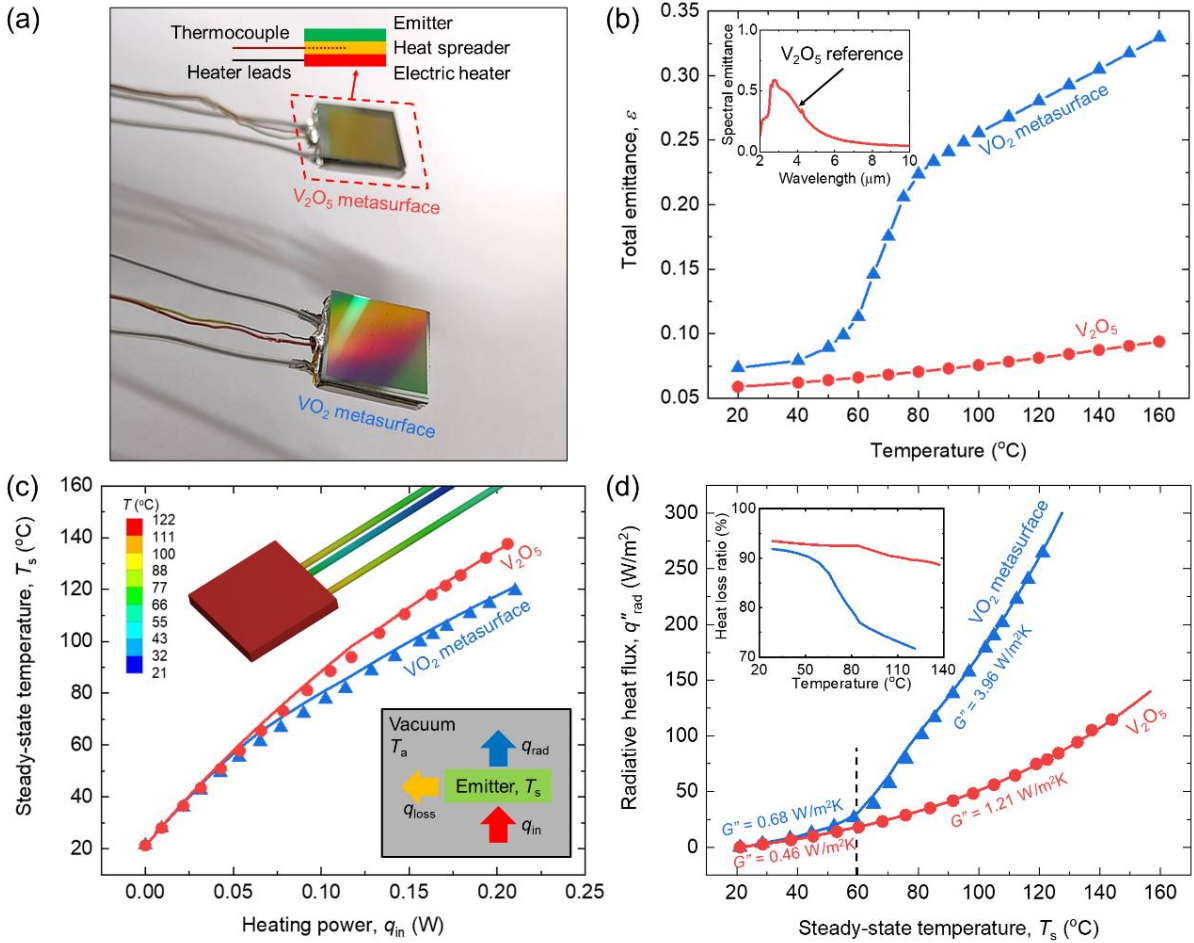

**Figure 4.** Thermal vacuum test for both tunable VO$_2$ and static V$_2$O$_5$ emitters. (a) Photo of samples and schematic of heating setup. (b) Total emittance as a function of temperature calculated based on measured temperature-dependent spectral emittance. Inset: measured temperature-independent spectral emittance of the V$_2$O$_5$ reference emitter. (c) Steady-state temperature at varying heating powers, where markers are experimental data and lines are simulated results. Top inset: simulated temperature distribution. Bottom inset: energy balance model. (d) Radiative heat flux emitted by the samples at varying steady-state temperature, where markers are processed experimental data by removing the parasitic heat loss and lines are simulated results. Inset: simulated parasitic heat loss as a function of sample temperature.

In order to validate the thermal measurement, thermal modeling was performed to numerically simulate the steady-state sample temperature at different heating power input (See Methods for details about thermal simulation). As plotted in Figure 4(c), the simulation results (solid lines) are in excellent agreement with the experimental data (markers) from the thermal vacuum test for both emitter samples. The simulated temperature distribution of the tunable $VO_2$ metasurface emitter at a heating power of 0.21 W is presented as the top inset. Clearly the heat is not only dissipated by thermal emission from the top emitter surface ($q_{rad}$), but also lost via conduction through electrical leads and thermocouple wires as well as radiation from the side and bottom surfaces ($q_{loss}$). As depicted in the bottom inset, energy balance at steady state yields $q_{in} = q_{rad} + q_{loss}$, where the parasitic heat loss $q_{loss}$ can be determined from the thermal simulation. The calculated $q_{loss}/q_{in}$ ratios for the tunable $VO_2$ metasurface and the static $V_2O_5$ emitters are provided at different steady-state temperatures in the inset of Figure 4(d), where the parasitic heat loss could be as high as 90% at low temperatures because of low emittance of the tunable metasurface with insulating $VO_2$ (and static $V_2O_5$ emitter), and reduced to 70% at higher temperature with increased emittance after $VO_2$ phase change.

By subtracting the parasitic heat loss $q_{loss}$ from experimental heating power $q_{in}$, the experimental radiative heat flux $q_{rad}$ from the thermal vacuum test for both the tunable $VO_2$ and static $V_2O_5$ metasurface emitters can be obtained at different sample temperatures $T_s$ displayed in Figure 4(d) as markers. When the temperature is below 60°C before the $VO_2$ phase change occurs, both emitter samples exhibit almost the same radiative heat flux increment with temperature quantified by a radiative thermal conductance (per unit area) value of $G'' = 0.68 \text{ W/m}^2\text{K}$ (from 20°C to 60°C) linearly fitted as $G'' = \partial q_{rad}/\partial T_s$. When the $VO_2$ transits into metallic phase that excites the MP with further increasing temperatures, the radiative heat flux increases drastically with the tunable VO$_2$ metasurface, characterized by a radiative thermal conductance value of $G'' = 3.96$ W/m$^2$K (fitted from 60°C to 120°C), clearly demonstrating 5.8 times enhancement in radiative heat dissipation. In particular, the radiative heat flux is increased by over 10 times from the phase change onset temperature 60°C ($q_{rad}$ = 26 W/m$^2$) to 120°C ($q_{rad}$ = 265 W/m$^2$) with the tunable metasurface emitter, thanks to the VO$_2$ phase change and more importantly, excitation of MP that leads to augmented spectral emittance at higher temperatures. By comparison, radiative heat flux of the static V$_2$O$_5$ reference emitter at 120°C (i.e., 75 W/m$^2$) is four times of that of itself at 60°C merely due to the red-shift of blackbody spectrum at higher temperatures, and is only 28% of that of the VO$_2$ counterpart at the same temperature. The radiative thermal conductance of the V$_2$O$_5$ emitter is $G'' = 1.21$ W/m$^2$K from 60°C to 150°C, only 1.6 times larger than that of itself at the lower temperatures because of the static spectral emittance, which strikingly highlights the superior performance of the tunable VO$_2$ metasurface emitter by 3.3 times (i.e., radiative thermal conductance ratio) for enhanced radiative heat dissipation. The experimental radiative heat fluxes from both emitters are also well validated by the excellent agreement with theoretical calculations as shown in Figure 4(d) as solid lines from $q_{rad} = \varepsilon\sigma(T_s^4 - T_a^4)$, where $\varepsilon$, $\sigma$, and $T_a$ are respectively the measured total emittance, Stefan-Boltzmann constant, and vacuum chamber wall temperature (21°C). Undoubtedly, the thermal vacuum tests have clearly demonstrated the enhanced radiative heat dissipation performance from the tunable metasurface by thermally excitation of MP with metallic VO$_2$ at high temperatures.

## Conclusions

In summary, we have successfully fabricated a tunable metasurface emitter made of VO$_2$ microdisks from an etch-free and scalable process. The measured spectral emittance of the tunable

VO$_2$ metasurface clearly exhibits total emittance enhanced over three times upon VO$_2$ phase change, in addition to the diffuse emission behavior with angular and polarization-independence. The augmentation of the infrared emittance mainly due to the switchable thermal emission peak at wavelength of $\lambda$ = 5 μm is thoroughly elucidated by thermally exciting or suppressing MP with either metallic or insulating VO$_2$, which is verified by analytical LC model along with electromagnetic field distribution. The enhanced radiative heat dissipation of the tunable VO$_2$ metasurface is directly demonstrated with the thermal vacuum test quantified by almost six-fold enhancement in radiative thermal conductance upon VO$_2$ phase transition and 3.3 times improvement over that of the static V$_2$O$_5$ emitter. Note that the radiative cooling performance can be optimized by tailoring the switchable MP emittance peak to best match the blackbody spectrum around the given operating temperatures by changing microdisk geometry.[28] The phase change temperature of VO$_2$ can be possibly lowered by doping with transition metals like tungsten[30] to better serve the applications around room temperatures. The results and understanding gained here will facilitate the applications of tunable metasurfaces in dynamic radiative heat dissipation and thermal management.

## Experimental section

***Sample fabrication***. Starting with a lightly doped Si wafer (resistivity > 20 Ω-cm, Virginia Semiconductor Inc.), a layer of 200-nm-thick aluminum was deposited using electron beam evaporation (Lesker PVD75 Electron Beam Evaporator) at a deposition rate of 1 Å/s and pressure of $5\times10^{-6}$ Torr. The 300-nm-thick HfO$_2$ was deposited by atomic layer deposition (Cambridge Savannah ALD Deposition Tool) with a Tetrakis(dimethylamino)hafnium (TDAHF) precursor at a growth rate of 1 Å per cycle. The photoresist (AZ 3312, AZ Electronic Materials) was spin coated on the HfO$_2$ spacer with a speed of 5000 rpm and exposed in a GCA 8500 Stepper with 5× size reduction from the patterns on a photomask (chrome mask on quartz substrate, Photo Science Inc.). Pure vanadium (99.99% purity, Kurt J. Lesker Co.) was deposited into the holes with a thickness of 50 nm by electron beam evaporation at a deposition rate of 0.8 Å/s and pressure of $5\times10^{-6}$ Torr. In the lift-off process, the residual photoresist and extra V were removed by soaking in acetone and sonication at room temperature for 1 minute. The sample was then transferred into a furnace (Mini-Brute tube furnace) at 300°C with a nitrogen flow rate of 60 SLPM and an oxygen rate of 0.5 SLPM for 8 hours. The heights of the V and $VO_2$ disks were measured by a profilometer (Dektak stylus profilers, Bruker Co.). The $V_2O_5$ reference emitter with the same geometric structure as the $VO_2$ metasurface was prepared by over-oxidizing V in the furnace for 24 hours.

***Optical characterization.*** The spectral properties were characterized by a Fourier-transform infrared spectrometer (Nicolet iS50, Thermal Fisher Scientific) and a variable-angle reflection accessory (Seagull, Harrick Scientific). The polarization of incident light was selected by a motorized infrared polarizer inside the spectrometer. A gold coated mirror was measured prior to the sample as a reference. The sample was mounted on a home-built heating stage, which consists of a copper heat spreader, a K-type thermocouple, an electric patch heater (KHLVA-101, Omega Engineering Inc.), and a temperature controller (CSi8DH, Omega Engineering Inc.). For each spectral measurement, 100 scans at a spectral resolution of 0.48 cm$^{-1}$ were taken. Both the sample and the reference were measured three times, and the average was reported. With the measured spectral reflectance $R_\lambda$, the spectral absorptance $\alpha_\lambda$ is obtained through $\alpha_\lambda = 1-R_\lambda$ as the sample is opaque at the wavelength range of interest. The spectral emittance $\varepsilon_\lambda$ is equal to $\alpha_\lambda$ according to Kirchhoff's law. The total emittance is computed as the ratio of the emissive power of a metasurface emitter to that of a blackbody at the same temperature over the wavelength range from 1 to 25 μm, i.e., $\varepsilon(T) = \frac{\int_{1\mu m}^{25\mu m} \varepsilon_\lambda(\lambda,T) E_{\lambda,b}(\lambda,T) d\lambda}{\int_{1\mu m}^{25\mu m} E_{\lambda,b}(T) d\lambda}$, where $\varepsilon_\lambda(\lambda, T)$ and $E_{\lambda,b}(T)$ are respectively the measured spectral emittance of the emitter and the spectral emissive power of a blackbody at temperature $T$ and wavelength $\lambda$.

***Optical modeling.*** In the finite-difference time-domain (FDTD) simulation (Lumerical Inc.), a domain area of 1.6×1.6×10 μm³ ($x \times y \times z$) was considered for one unit cell of $VO_2$ microdisk arrays with a disk diameter of 1.1 μm. The boundary conditions (BC) along the $x$ and $y$ directions were set as periodic, and that along the $z$ direction was perfectly matched layer (PML). An auto non-uniform mesh was employed, and a convergence check was conducted to obtain mesh-independent results. The optical properties of the Al substrate and $HfO_2$ spacer are taken from the literature,[31-32] while those of insulating and metallic $VO_2$ were fitted from our previous measurement.[33] The spectral reflectivity under normal incidence was simulated, and the emittance was calculated through $\varepsilon_\lambda = \alpha_\lambda = 1 - R_\lambda$.

***Thermal vacuum test***. An emitter sample of 15×15 mm² was mounted on a copper heat spreader, which was heated by a ceramic heater. A K-type thermocouple was inserted into the copper heat spreader from the side for temperature measurements along with a data acquisition module (OM-USB-TEMP, Omega Engineering Inc.). The thermal test was conducted inside a 24-inch bell jar vacuum chamber at a pressure lower than 5×10⁻³ Torr, at which conduction and convection heat transfer by air can be safely neglected. The heating power was increased incrementally by a source meter (Keithley 2400), and sample temperature was recorded after the steady state was reached at a given heating power. The accuracy for the temperature and power measurements are ±1.5°C and ±0.13%, respectively.

***Thermal modeling.*** In ANSYS software, a model including the sample, heat spreader, heater, thermocouple, and leads was developed using the build-in DesignModeler. Corresponding materials were assigned to the components, e.g., silicon to the sample, copper alloy to the spreader, etc. Conformal meshes were generated, and the mesh size was carefully checked to obtain mesh-independent results. Boundary conditions were set as follows. Constant heat flux was given to the heater, and radiation was assigned to all exposed surfaces with ambient temperature of 21°C and measured thermal emittance. Properties of each component, including material, thermal conductivity, geometry, and total emittance, are summarized in Table 1. The end sides of the thermocouple and heater leads were set as thermally insulated, and convection was ignored in vacuum. In the Steady-State Thermal module, the power of the heater, which was measured by the source meter, and the measured emittance of the sample were input, and the steady-state temperatures and heat flux at each surface were computed. In ANSYS, these heat losses were determined by inserting heat flux probes at the surfaces of interest.

Table 1 Properties used in ANSYS of each component.

| Component | Material | Thermal conductivity [W/mK] | Geometry [mm] | Total emittance |
|---|---|---|---|---|
| Sample | Silicon | 124 | 15×15×0.5 | Figure 4(b) |
| Heat spreader | Copper alloy | 401 | 15×15×0.5 | 0.15 (Al tape) |
| Heater | Alumina | 24 | 10×10×1.2 | 0.15 (Al tape) |
| Thermocouple, insulation | Teflon | 0.25 | 0.1×0.45×100 | 0.5 |
| Thermocouple, wire | Nickel-cobalt-chromium alloy | 11 | 0.8×0.25×100 | N/A (not exposed) |
| Heater lead, insulation | Teflon | 0.25 | 1×0.25×100 | 0.5 |
| Heater lead, wire | Nickel | 69.9 | 0.5×0.5×100 | N/A (not exposed) |

## AUTHOR INFORMATION

**Corresponding Author**

* Liping Wang - School for Engineering of Matter, Transport & Energy, Arizona State University, Tempe, Arizona, USA 85287; Email: liping.wang@asu.edu

**Authors**

Linshuang Long - School for Engineering of Matter, Transport & Energy, Arizona State University, Tempe, Arizona, USA 85287; Email: linshuang.long@asu.edu

Sydney Taylor - School for Engineering of Matter, Transport & Energy, Arizona State University, Tempe, Arizona, USA 85287; Email: sjtayl12@asu.edu

**Author Contributions**

L.L. fabricated and characterized (profilometer, Raman, FTIR) the samples as well as conducted optical modeling, thermal vacuum test, and thermal modeling. S.T. performed additional sample characterization (SEM). L.W. supervised the overall project. The initial manuscript was written by L.L. and revised by L.W. All authors have reviewed and approved the final version of the manuscript.

**Notes**

The authors declare no competing financial interest.

## ACKNOWLEDGEMENTS

This work was mainly supported by National Science Foundation (NSF) under Grant No. CBET-1454698. S.T. would like to thank support from NASA Space Technology Research Fellowship (NNX16AM63H). Sample fabrication and characterization was supported by ASU NanoFab and Eyring Center under NSF contract ECCS-1542160.